%% file: main.tex
\documentclass[conference]{IEEEtran}
\IEEEoverridecommandlockouts
\usepackage[pdftex]{graphicx}
\graphicspath{{figs/}{/figs/}}
\DeclareGraphicsExtensions{.pdf,.jpeg,.png}

\usepackage{url}
\usepackage[dvipsnames]{xcolor}
\usepackage[para]{threeparttable}
\usepackage{array}

\usepackage[utf8]{inputenc}
\usepackage[T1]{fontenc}
\usepackage{bm} \usepackage{amsmath,times,booktabs, tabularx, hyperref, subcaption, multirow}
\usepackage{amssymb, amsfonts}
\usepackage{microtype}

\usepackage[backend=biber,                          style=ieee,                             ]{biblatex}
\usepackage{algorithmic}
\usepackage{textcomp}

\def\BibTeX{{\rm B\kern-.05em{\sc i\kern-.025em b}\kern-.08em
    T\kern-.1667em\lower.7ex\hbox{E}\kern-.125emX}}

\usepackage[capitalise, nameinlink]{cleveref}

\usepackage{xcolor}

\usepackage{siunitx}
\usepackage[automake, nomain, acronym, xindy, shortcuts=ac]{glossaries-extra}
\glsdisablehyper
\makeglossaries
\setabbreviationstyle[acronym]{long-short} \glssetcategoryattribute{acronym}{nohyperfirst}{true}
\loadglsentries{glossary}

\newcommand{\dent}[0]{\raisebox{1mm}{~~$\llcorner$~}}

\newcommand{\emailaddr}[1]{{\small\normalfont\texttt{#1}}}
\newcommand{\orcidauthor}[2]{{\em {#1}}\hspace*{1mm}\href{https://orcid.org/#2}{\includegraphics[width=1em]{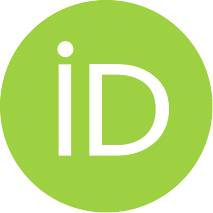}}}

\begin{document}
\title{WhaleVAD-BPN\\\large Improving Baleen Whale Call Detection with Boundary Proposal Networks and \\Post-processing Optimisation}

\author{\IEEEauthorblockN{\orcidauthor{Christiaan M. Geldenhuys}{0000-0003-0691-0235}}
\IEEEauthorblockA{
\emailaddr{cmgeldenhuys@sun.ac.za}
}
\and
\IEEEauthorblockN{\orcidauthor{G\"{u}nther Tonitz}{0009-0009-6030-4122}}
\IEEEauthorblockA{
\emailaddr{gtonitz@sun.ac.za} \\
\\
    \textit{Department of Electrical and Electronic Engineering}\\
    \textit{University of Stellenbosch}\\
    Stellenbosch, South Africa
}
\and
\IEEEauthorblockN{\orcidauthor{Thomas R. Niesler}{0000-0002-7341-1017}}
\IEEEauthorblockA{
\emailaddr{trn@sun.ac.za}
}
}

\newcolumntype{P}[1]{>{\centering\arraybackslash}p{#1}}

\maketitle

\begin{abstract}
While recent sound event detection (SED) systems can identify baleen whale calls in marine audio, challenges related to false positive and minority-class detection persist.
We propose the boundary proposal network~(BPN), which extends an existing lightweight SED system.
The BPN is inspired by work in image object detection and aims to reduce the number of false positive detections.
It achieves this by using intermediate latent representations computed within the backbone classification model to gate the final output.
When added to an existing SED system, the BPN achieves a 16.8\,\% absolute increase in precision, as well as 21.3\,\% and 9.4\,\% improvements in the F1-score for minority-class \texttt{d}-calls and \texttt{bp}-calls, respectively.
We further consider two approaches to the selection of post-processing hyperparameters: a forward-search and a backward-search.
By separately optimising event-level and frame-level hyperparameters, these two approaches lead to considerable performance improvements over parameters selected using empirical methods.
The complete \gls{wadbpn} system achieves a cross-validated development F1-score of 0.475, which is a 9.8\,\% absolute improvement over the baseline.
\end{abstract}

\vspace*{1mm}

\section{Introduction}
\Ac{pam} has become a cornerstone for assessing marine mammal populations in remote and otherwise inaccessible habitats, owing to its non-invasive nature and relatively modest financial cost.
In practice, however, \ac{pam} generates substantial volumes of data, often suffering from low signal-to-noise ratios in recordings~\autocite{Kowarski2021BaleenReview}.
Both factors render manual annotation labour-intensive and dependent on specialist knowledge.
Consequently, a large body of research has focused on the development of automated detectors for the vocalisations of key species, most notably \textit{Balaenoptera~musculus}~(blue) and \textit{B.~physalus}~(fin) whales.
These species remain classified as endangered and vulnerable by the IUCN~\autocite{Cooke2018FinWhale,Cooke2018BlueWhale}, and continue to elude precise abundance estimates because of a scarcity of labelled data~\autocite{miller2021atbfljournal}.

Recent \ac{sed} systems have made considerable strides in identifying baleen whale calls within marine recordings, yet these models still exhibit a high false-positive rate and do not allow the reliable detection of infrequent~(minority-class) call events~\autocite{Geldenhuys2025WhaleVAD}.
To address these shortcomings, we propose the inclusion of a complementary neural module, the \ac{bpn}.
Inspired by the field of object detection in image processing~\autocite{Ren2015FasterRCNN, Liu2016SSD}, the \ac{bpn} is intended to supplement an existing lightweight \ac{sed} architecture by exploiting intermediate latent features already calculated within the backbone classifier.
The \ac{bpn} output is used as a gating mechanism that refines the temporal localisation of detected events, thereby reducing false positives and improving overall precision.
It was found that the \ac{bpn} also leads to an improvement in recall for minority-class call types.

In addition to the described architectural augmentation, we introduce improvements to the post-processing stage of the \ac{sed} architecture itself.
We investigate two search strategies for optimising the hyperparameter selection of the post-processing stage: a forward-search and a backward-search.
These procedures systematically explore event- and frame-level hyperparameters, yielding performance gains that surpass those obtained through existing ad hoc or empirical methods.

\section{Background}
This section provides a summary of the background regarding baleen whale call detection, proposal networks, and the post-processing techniques used.

\subsection{Baleen whale call activity detection}
Baleen whales~(\textit{mysticetes}) undertake extensive migrations and thus communicate using low-frequency vocalisations that propagate over great distances underwater.
Previous work has shown that \ac{vad} algorithms can be applied to baleen whale call detection, specifically detecting the calls of blue and fin whales~\autocite{Geldenhuys2025WhaleVAD}.
For example, the AVA-VAD~\autocite{Wilkinson2021AVAVAD} system relied on producing latent features from a spectrogram representation of the audio recording, using a \ac{cnn}.
These latent features are then processed sequentially by a \ac{bilstm} architecture with sigmoid activation to obtain the final posterior classification probabilities for each frame in the input spectrogram.

While phase information has been disregarded in speech processing systems, its inclusion
has been reported to afford a \qty{10}{\percent} improvement in the F1-score~\autocite{Geldenhuys2025WhaleVAD}.
This work proposed further changes to the AVA-VAD architecture, with the addition of bottleneck and depthwise convolutional layers with recurrent connections.
The final proposed system was called \gls{wad}, and it is this system that we will extend.

\subsection{Proposal networks}
Proposal networks are a two-stage architecture common in the computer vision field for achieving object detection~\autocite{Girshick2014RCNN,Girshick2015FastRCNN,Ren2015FasterRCNN}.
These networks have been refined over time to be computationally efficient and have proven to be effective at localising and identifying objects within an image.
Early approaches, such as R-CNN~(regions with \acs{cnn} features)~\autocite{Girshick2014RCNN}, relied on external algorithms, such as \textit{selective search}~\autocite{Uijlings2013SelectiveSearch}, to generate a set of candidate \acp{roi}.
These \acp{roi} could then be used by a secondary detection network to determine if a particular object is present in the proposed region.
However, these proposal networks relied on handcrafted convolutional features.

\Textcite{Ren2015FasterRCNN} were the first to introduce the \ac{rpn}, which integrated proposal generation directly into the neural network architecture.
The technique relied on applying a small trainable \ac{cnn} to the feature map of a convolutional backbone network to coordinate and refine a set of predefined anchor boxes.
By sharing convolutional features with the downstream detection network, the \ac{rpn} enabled an efficient and end-to-end trainable system.
This architecture established the foundation for most subsequent two-stage object detectors, demonstrating the effectiveness of backbone features.

Despite the success of \acp{rpn}, its reliance on a single, deep feature map presents limitations in detecting objects across varying scales and remains computationally less efficient when compared to single-stage detectors.
\Textcite{Liu2016SSD} address this by attaching multiple detection heads to intermediate layers at varying depths within the backbone network.
Other single-stage detectors, such as YOLOv3~\autocite{Redmon2018YOLOv3}, adopted a multi-scale \ac{fpn}[-inspired]~\autocite{Lin2017FeaturePyramidNetwork} mechanism to aggregate intermediate features.

\subsection{Post-processing}
\label{subsec:background_post_processing}

During per-frame \ac{sed} in a bioacoustic system, the classification model produces a sequence of probabilities over time, each indicating the likelihood of a particular sound class being present in a frame.
These \emph{model probabilities} are converted to \emph{binary detections} by applying a threshold to each of the per-frame call probabilities.
These binary detections are then aggregated into \emph{discrete events} representing contiguous periods of~(call)~activity.

The ideal system output is a single event that accurately aligns with the human annotation.
In practice, however, these output events are often fragmented, containing intermittent gaps, or spurious or excessively long detections.

To mitigate this, filtering is applied during post-processing.
We consider two classes of filtering techniques, namely frame- and event-level.

\subsubsection{Frame-level techniques}
\label{subsubsec:frame_level_techniques}

These methods serve to smooth either the model output probabilities, or the detections, or both.
For example, a median filter can be applied to the model output, where each probability is replaced by the median value within a neighbourhood.
This reduces sporadic peaks or dips in activity, thereby decreasing the number of fragmented events, at the cost of reduced precision.
Hysteresis is another common frame-level post-processing approach, where a window of past estimates influences the current model decision.
One implementation applies different thresholds for entering and exiting an event state, which we will refer to as \textit{threshold hysteresis}~\autocite{Cances2019EvalPostprocessingDCASE}.
By setting a lower threshold for termination, the system remains active for longer, which may help reduce event fragmentation.
Alternatively, activity at a given time instant can be defined based on the majority vote~(statistical mode) within the sliding window, which we will refer to as \textit{hangover}.
This can be interpreted as a variant of median filtering applied to model binary detections, as opposed to model class probabilities; given by the following equation:
\begin{equation}
    \tilde{y}_t =
    \begin{cases}
        1 & \text{if } \displaystyle \sum_{i=0}^{k} \hat{y}_{t-i} > \frac{k+1}{2}, \\[8pt]
        0 & \text{otherwise},
    \end{cases}
\end{equation}
where $\tilde{y}_t$ is the model detection at time instant $t$, and $k$ is the number of past samples in the sliding window.

\subsubsection{Event-level techniques}
\label{subsubsec:event_level_techniques}

These methods act at an event-level on the aggregated detections and commonly impose constraints on duration.
Minimum and maximum event durations may be defined to remove events of implausible duration, and a minimum inter-event duration may be defined to merge events occurring close in time.

The effectiveness of these post-processing methods depends on a set of hyperparameters, such as the size of the neighbourhood used to implement median filtering or hangover.
The selection of these hyperparameters is highly task and data-specific, and thus requires optimisation.

\section{Literature Review}
This section starts with an overview of previous research in the detection and classification of baleen whales, followed by a focus on post-processing strategies that have been implemented and evaluated in bioacoustic \ac{sed}.

\subsection{Detection and classification of baleen whale calls}

Automated methods for the detection and classification of baleen whales continue to develop.
Due to the great distances over which their vocalisations propagate underwater, these calls are an ideal candidate for \ac{pam}~\autocite{Mellinger1997MysticeteMethods}.

Before the advent of machine learning, automated detection relied on classical signal-processing techniques designed to identify signals with known, empirical characteristics.
Two common approaches were matched filtering, in which a synthetic kernel derived from a known call is correlated with a recording to locate vocalisations in background noise~\autocite{Mellinger1997MysticeteMethods}, and spectrogram correlation, which cross-correlates a template spectrogram with successive recording segments to identify vocal occurrences~\autocite{Mellinger2000SpecCor}.

Machine learning approaches were subsequently adopted to address the non-stationary nature of ocean soundscapes and reduce the reliance on fixed filters or templates.
Examples include \acp{svm} to identify individual humpback whale vocalisations~\autocite{Mazhar2007SVMHumpback} and North Atlantic right whale upcalls~\autocite{Ibrahim2016SVMNorth}, as well as probabilistic models such as \acp{gmm} for blue whale calls~\autocite{Cuevas2017GMMBlue} and \acp{hmm} for humpback whale calls~\autocite{pace2012HMMHumpback}.

More recently, \acp{dnn}, particularly \acp{cnn}, have become a common area of research for computational bioacoustics.
Large \ac{cnn} architectures, such as DenseNet~\autocite{Huang2018DenseNet}, have been applied to baleen whale classification; for example, \textcite{Miller2023dcalls} reported that a DenseNet-based system surpassed human observers in blue whale D-call detections.
\Acp{crnn}, which integrate recurrent layers to capture temporal dependencies, have also shown strong performance in fin- and blue whale classification~\autocite{Rasmussen2021AutomaticDA}.
Despite continued improvements in frame-level detector performance~\autocite{Schall2024Baleen}, lower false-positive rates are desirable, as frame-level errors can cascade into greater event-level errors.

\subsection{Post-processing in bioacoustic SED}

Frame-level post-processing commonly involves class-dependent thresholds~\autocite{hoffman2024DCASE, Zhao2024FewShotBE, miller2021atbfljournal} and median filters with a fixed kernel size~\autocite{Zhao2024FewShotBE}.
\Textcite{hoffman2024DCASE} report that class-specific thresholds, applied to logits, can outperform a fixed threshold on bird-call detection.
Hysteresis thresholds have not been reported in the bioacoustic literature.
However, this technique has been applied in other \ac{sed} domains.
For example, \textcite{Cances2019MULTITASKLA} found hysteresis thresholds to outperform absolute thresholds when classifying sound classes that occur in a domestic environment.
Hangover has also not been reported on in \ac{sed} studies, although it has been applied successfully in other detection problems, such as sonar~\cite{Barshan1998PerformanceCO} and radar~\cite{difranco1980radar}, where an \textit{M-out-of-N} criterion is used.

Event-level post-processing techniques are common in few-shot bioacoustic learning.
For example, several studies impose a minimum event duration and a minimum inter-event interval.
These hyperparameters are commonly derived from the support set~\autocite{hoffman2024DCASE, Liu2022SurreySF}.
\Citeauthor*{miller2021atbfljournal}~\autocite{miller2021atbfljournal} apply a minimum and maximum event duration of \qty{0.5}{s} and \qty{2.5}{s} respectively, and require at least \qty{0.5}{s} between events when identifying fin whale \qty{20}{\hertz} pulses.

For domestic environmental \ac{sed}, \textcite{Cances2019EvalPostprocessingDCASE} have provided a comprehensive comparison of frame-level post-processing techniques, including fixed threshold, hysteresis threshold, and a slope-based threshold technique which detects fast changes in model probabilities.
Each of the threshold techniques were evaluated using class-dependent and class-independent hyperparameters.
The authors report an \qty{28.6}{\percent} absolute improvement in the F1-score over the baseline, when selecting class-dependent fixed thresholds.
This suggests that the selection of class-specific hyperparameters and the use of multiple post-processing strategies in bioacoustic \ac{sed} may yield substantial gains in detection performance.
However, to our knowledge, no bioacoustic \ac{sed} study has reported a systematic treatment of different post-processing strategies.

\begin{table*}[h!]
    \centering
    \caption{Summary of the \acs{atbfl} site-year training and development sets.
    The table shows the average recording duration~(hours), number of recordings, total recording duration~(hours), number of annotated events, and total duration of whale calls~(hours) for each set.
    }
    \label{tbl:recording_info}
    \begin{tabular}{@{}lccccc@{}}
        \toprule
        Dataset  & Avg. Duration (h) & Recordings & Total duration~(h) & Total events & Total event duration~(h) \\
        \midrule
        \texttt{ballenyisland2015}           & 1.0  & 205  & 204  & 2222  & 2.8  \\
        \texttt{casey2014}                   & 1.0  & 194  & 194  & 6866  & 14.2 \\
        \texttt{elephantislands2013}         & 0.08 & 2247 & 187  & 21949 & 16.1 \\
        \texttt{elephantislands2014}         & 0.08 & 2595 & 216  & 20962 & 28.1 \\
        \texttt{greenwich2015}               & 0.17 & 190  & 32   & 1128  & 2.1  \\
        \texttt{kerguelen2005}               & 1.0  & 200  & 200  & 2960  & 3.5  \\
        \texttt{maudrise2014}                & 0.42 & 200  & 83   & 2360  & 5.7  \\
        \texttt{rosssea2014}                 & 1.0  & 176  & 176  & 104   & 0.1  \\
        \midrule \textbf{Total training set}          & -- & \textbf{6007} & \textbf{1292} & \textbf{58551} & \textbf{72.6} \\
        \midrule \texttt{casey2017}                   & 1.0  & 187  & 185  & 3263  & 6.1  \\
        \texttt{kerguelen2014}               & 1.0  & 200  & 200  & 8822  & 11.4 \\
        \texttt{kerguelen2015}               & 1.0  & 200  & 200  & 5542  & 7.4  \\
        \midrule \textbf{Total development set}       & -- & \textbf{587} & \textbf{585}  & \textbf{17627} & \textbf{24.9} \\
        \bottomrule
    \end{tabular}
\end{table*}

\section{Data}

\begin{figure}[h!]
  \centering
  \includegraphics[width=0.9\linewidth]{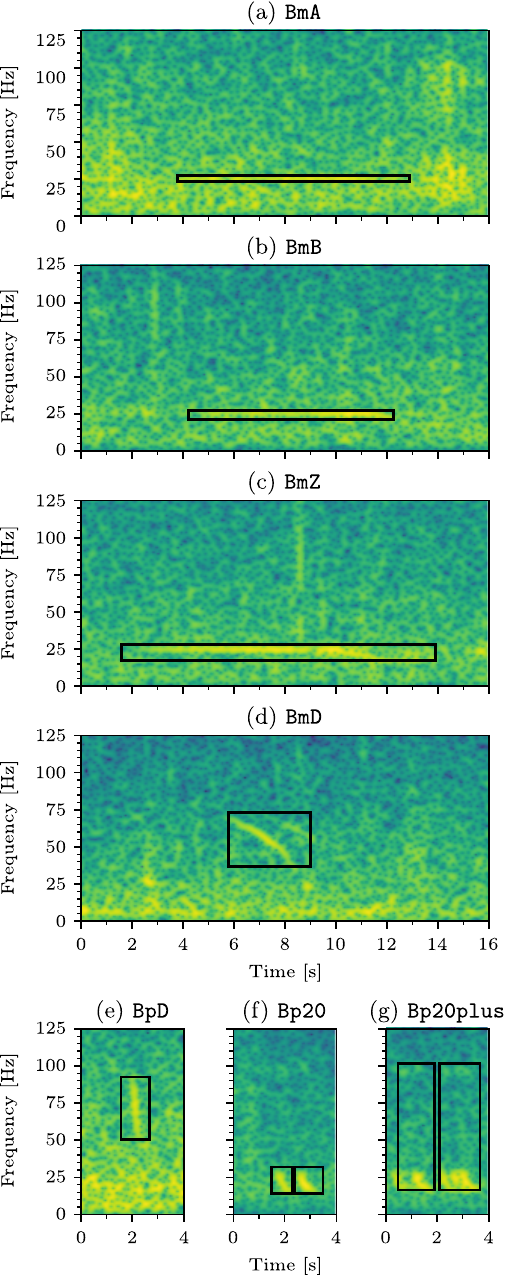}
  \caption{
  Spectrogram representations of exemplar baleen whale call types~\autocite{JeanLabadye2025BioDCASE2025Task2DevDataset}.
  Bounding boxes indicate presence of particular call type, provided by domain expert annotations in dataset.
  Figures (a)\,–\,(d) correspond to blue whale vocalisations, and (e)\,–\,(g) correspond to fin whale vocalisations.
  }
  \label{fig:call_examples}
\end{figure}

As part of the 2025 IEEE BioDCASE~(Task 2) challenge, a dataset consisting of strongly labelled blue- and fin-whale calls in the Antarctic region of the Southern Ocean was released~\autocite{JeanLabadye2025BioDCASE2025Task2DevDataset}.
The data was originally obtained by the \ac{atp} as part of the \ac{iwc-sorp}~\autocite{Miller2020ATBFL}.

The \ac{atbfl} consists of 11 site-year datasets recorded around the Antarctic, in the period 2005 to 2017.
Sites were selected based on the availability of a full year of data, and utilised different recording instrumentation.
All audio data was resampled to a rate of \qty{250}{\hertz}.
Each dataset was manually annotated, in both the time and frequency domains, by domain experts using the data collection and annotation procedures described in \citeauthor*{miller2021atbfljournal}~\autocite{miller2021atbfljournal}.

The challenge identifies three site-year datasets as a development set, while the remaining eight site-year sets form the training set.
The entire dataset contains a total of \qty{6594}{recordings} with a total duration of \qty{76178}{hours}, as set out in \Cref{tbl:recording_info}.
Only \qty{5.2}{\percent} of the data~(in duration) contains blue or fin whale vocalisations, however.

The \ac{atbfl} includes annotations for seven different call types, of which four are produced by blue whales~(\textit{BmA}, \textit{BmB}, \textit{BmZ} and \textit{BmD}) and three by fin whales~(\textit{BpD}, \textit{Bp20} and \textit{Bp20plus}).
A blue whale Z-call~(\textit{BmZ}) is a low-frequency compounded call consisting of both the A-call~(\textit{BmA}) and the B-call~(\textit{BmB}).
The blue whale D-call~(\textit{BmD}) is a downsweeping call, ranging between \qtyrange{20}{120}{\hertz}, which is similar to a fin whale downsweep~(\textit{BpD}) which ranges from \qtyrange{30}{90}{\hertz}. Finally, fin whales also create a downsweeping pulse between \qtyrange{15}{30}{\hertz} which can either appear with~(\textit{Bp20plus}) or without~(\textit{Bp20}) an overtone varying between \qtyrange{80}{120}{\hertz}.
\Cref{fig:call_examples} shows examples of each call type.

Following \textcite{Schall2024Baleen}, call types are grouped based on their acoustic similarity and interrelated usage.
The A-, B-, and Z-calls, which co-occur, are combined into a single ABZ-call category~(\texttt{bmabz}).
Similarly, the blue whale D-call and the fin whale BpD-call are grouped into a unified D-call category~(\texttt{d}), while the Bp20 and Bp20Plus calls are merged into the Bp-call category~(\texttt{bp}).
These call grouped categories serve as the final call labels used.

\Cref{tbl:call_info} summarises the duration, frequency characteristics, and annotation counts for all call types found in the training and the development sets.
Blue whale vocalisations are more common in both datasets, comprising \qty{74.8}{\percent} of the training set and \qty{65.9}{\percent} of the development set annotations.
Notably, blue whale \mbox{A-calls,} \mbox{B-calls,} and Z-calls exhibit longer call durations, accounting for \qty{45.18}{hours} in the training set and \qty{20.9}{hours} in the development set.
This distribution reveals an imbalance, both in species representation (blue whales are overrepresented) and in the temporal occurrence of call types.

\begin{table}[t]
    \centering
    \caption{
    Different whale call frequency~(hertz) and duration~(seconds) information computed from data in~\autocite{JeanLabadye2025BioDCASE2025Task2DevDataset}.
    }
    \begin{subtable}{\linewidth}
        \centering
        \caption{Training Set}
        \begin{tabular}{@{}lccccccc@{}}
        \toprule
        \multirow{2}{*}{Type} & \multicolumn{3}{c}{Frequency (Hz)} & \multicolumn{3}{c}{Duration (s)} & \multirow{2}{*}{Count} \\
                    & Min   & Max   & Avg. & Min  & Max    & Avg.  & \\
        \midrule
        \texttt{BmA}         & 11.4  & 110.6 & 25.9 & 2.12 & 27.11  & 7.19  & 13785 \\
        \texttt{BmB}         & 10.0  & 31.3  & 22.2 & 3.14 & 19.51  & 7.83  & 5433  \\
        \texttt{BmZ}         & 11.5  & 34.6  & 22.0 & 3.87 & 28.07  & 12.76 & 1646  \\
        \texttt{BmD}         & 11.5  & 110.7 & 69.9 & 0.29 & 6.78   & 1.42  & 22977 \\
        \texttt{BpD}         & 16.7  & 134.1 & 75.4 & 0.29 & 2.70   & 1.12  & 2658  \\
        \texttt{Bp20}        & 8.5   & 45.1  & 22.2 & 0.48 & 3.08   & 1.52  & 9104  \\
        \texttt{Bp20plus}    & 9.2   & 112.7 & 52.5 & 0.76 & 2.91   & 1.50  & 3950  \\
        \bottomrule
        \end{tabular}
        \label{subtbl:call_info_train}
    \end{subtable}

    \vspace{1em}

    \begin{subtable}{\linewidth}
        \centering
        \caption{Development Set}
        \begin{tabular}{@{}lccccccc@{}}
        \toprule
        \multirow{2}{*}{Type} & \multicolumn{3}{c}{Frequency (Hz)} & \multicolumn{3}{c}{Duration (s)} & \multirow{2}{*}{Count} \\
                    & Min   & Max   & Avg. & Min  & Max    & Avg.  & \\
        \midrule
        \texttt{BmA}         & 15.7  & 30.1  & 25.6 & 2.12 & 36.62  & 7.12  & 6268  \\
        \texttt{BmB}         & 10.7  & 99.0  & 22.6 & 1.29 & 18.1   & 8.35  & 2277  \\
        \texttt{BmZ}         & 12.1  & 30.3  & 21.9 & 5.15 & 29.45  & 12.64 & 918   \\
        \texttt{BmD}         & 15.9  & 122.9 & 57.4 & 0.74 & 7.36   & 2.87  & 2168  \\
        \texttt{BpD}         & 26.5  & 137.5 & 61.7 & 0.37 & 2.58   & 1.08  & 688   \\
        \texttt{Bp20}        & 10.3  & 47.9  & 22.7 & 0.46 & 2.83   & 1.35  & 2550  \\
        \texttt{Bp20plus}    & 11.1  & 106.6 & 57.1 & 0.64 & 2.58   & 1.43  & 2758  \\
        \bottomrule
        \end{tabular}
        \label{subtbl:call_info_val}
    \end{subtable}

    \label{tbl:call_info}
\end{table}

\begin{figure*}[h]
    \centering
    \includegraphics[width=0.85\textwidth]{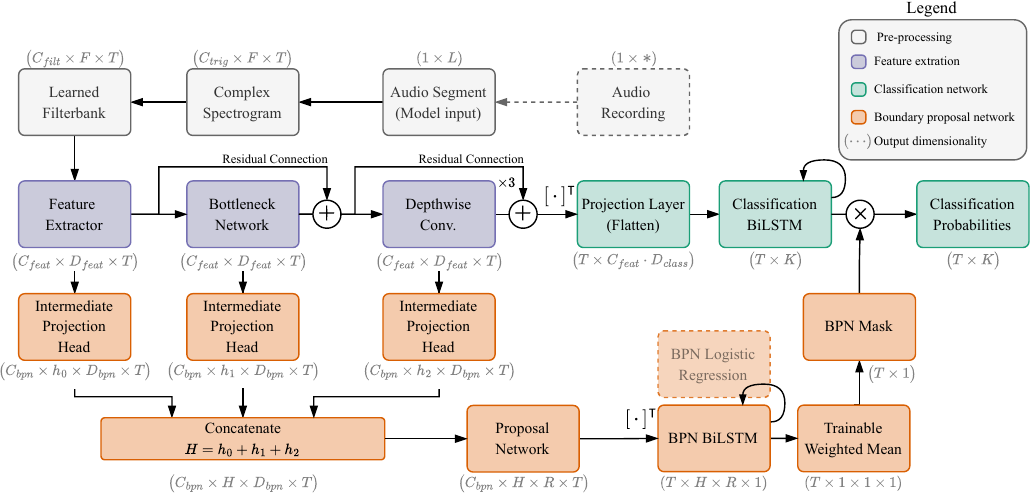}
    \caption{
    Illustration of the \gls{wadbpn} system we propose for improved whale call boundary detection.
    The system is divided into four sections: pre-processing, feature extraction, classification, and boundary proposal.
    The output tensor dimensionality is shown below each module in the system.
}
    \label{fig:wad-bpn}
\end{figure*}
\section{Experimental Structure}

In the following, we first describe the architectural changes we have made to the \gls{wad} system.
Then, we present our \acl{bpn},  which used intermediate features from within the \gls{wad} system to improve overall model performance.
Finally, we present two post-processing optimisation search strategies that separately optimise frame-level and event-level hyperparameters.

\subsection{Architecture modifications}
In the original \gls{wad} model~\autocite{Geldenhuys2025WhaleVAD}, the \textit{depthwise convolution} block consisted of three depthwise convolution layers placed in series.
We adapt the depthwise convolution block by adding residual connections between each of the layers with increasing dilation of \numlist{2;4;8}.
The increase in dilation factor provides a wider receptive field, thus allowing the network to utilise features that are further away in time~\autocite{Luo2016RecpetiveFieldCNN}.
Each depthwise convolution block retains the GELU activation and batch normalisation used by the original model.
In addition, the conventional dropout has been replaced with spatial dropout~\autocite{Thompson2015SpatialDropout}, which has been shown to improve the effective regularisation for convolutional layers.

\subsection{Boundary proposal network}

While the original \gls{wad} model architecture achieved high recall, the model also exhibited a high \ac{fpr}, which in turn resulted in a low precision~\autocite{Geldenhuys2025WhaleVAD}.
We propose a new component, the \acf{bpn}, whose purpose is to compute a gating score that is combined with the \gls{wad} classifier output in an effort to reduce false positives.
The \acf{bpn} uses convolutional feature representations obtained from intermediate layers, referred to as \textit{intermediate feature maps}, from the backbone \gls{wad} classifier, to compute this gating score.

The following subsections describe the components that comprise the \ac{bpn}, as well as the training regime used.

\subsubsection{Intermediate projection head}
Each set of intermediate feature maps is processed by a separate \ac{cnn}, called an intermediate projection head.
This consists of a convolutional layer with batch normalisation, GELU activation and a final maximum pooling layer.
Each projection head shares the same architecture, but has its own set of weights associated with a particular point in the backbone network, from which the feature map was drawn.

\subsubsection{Proposal network}
The outputs of all heads are concatenated along a new dimension~$H$, which is processed by the proposal network to produce $R$ distinct \ac{roi} vectors per head.
Each \ac{roi} consists of a latent feature vector with dimensionality~$C_{bpn}$, associated with a projection head~($h_i \in H$) for each time instant~$T$.
The proposal network consists of two convolutional layers with GELU activation, batch normalisation and spatial dropout~\autocite{Thompson2015SpatialDropout}.
We evaluate two variants of the network.
\emph{\ac{bpn}[-multi]} produces multiple \acp{roi} per projection head using a convolutional transpose~($R > 1$), whilst \emph{\ac{bpn}[-single]} produces a single \ac{roi} per projection head~($R = 1$).
Early experimentation showed that \ac{bpn}[-multi] outperformed \ac{bpn}[-single] and thus only \ac{bpn}[-multi] was considered in the final results, and will be referred to as simply \ac{bpn} throughout.
\Cref{tbl:wadbpn-config} shows a summary of the \ac{bpn} layer configuration.

\subsubsection{\Acs{bilstm}}
Each latent \ac{roi} vector is processed sequentially in time by a \ac{bilstm} or independently by a \ac{lr} module; with a sigmoid activation on the output.
During hyperparameter optimisation, it was found that the \ac{bilstm} network outperforms \ac{lr}.
The resulting output is averaged over each \ac{roi} using a learned weighted mean to produce a \textit{mask}.
The weighted average is jointly trained with the model, resulting in some heads being weighted more heavily than others.
This weighting remains fixed during inference.

\subsubsection{Masking}
The final mask is applied to the posterior call probabilities produced by the backbone classifier, thus acting as a soft gating mechanism whose purpose is to suppress spurious detections~(false positives).
As a result, the final posterior call probabilities are dependent on both the classification network, which is responsible for localising and identifying a particular call type in time, and the gating mechanism of the \ac{bpn} which aligns with this postulated call.
The addition of the \ac{bpn} gating mechanism should therefore allow the number of false positive classifications made by the original \gls{wad} architecture to be reduced through training.
We will refer to our model as \gls{wadbpn}.
\Cref{fig:wad-bpn} shows an illustration of the complete \gls{wadbpn} system.

\subsubsection{Training regime}
All models were trained using the AdamW~\autocite{losh2017adamw} optimiser with Focal loss~\autocite{Lin2018FocalLoss}.
The training set was divided into mini-batches of \qty{48}{segments} per batch, each consisting of approximately \qty{30}{seconds} long.
The learning rate is kept fixed at \num{0.001} with momentum terms of \numlist{0.9; 0.999} and a weight decay factor of \num{0.01}.
Training is halted once the training loss has converged or after \qty{32}{epochs} over the entire training set.

\begin{table}[t]
\centering
\caption{
    The \acf{bpn} layerwise configuration in terms of kernel size~$(K)$, stride~$(S)$, number of input channels~$\left(C_{in}\right)$ and output channels~$\left(C_{out}\right)$.
}
\label{tbl:wadbpn-config}
\begin{tabular}{@{}llcccc@{}}
\toprule
Layer &  & $K$ & $S$ & $C_{in}$ & $C_{out}$ \\ \midrule
Intermediate projection head \\
\dent Conv2D && (1, 1) & (1, 1) & 128 & 128 \\
\dent Max pool && (3, 1) & (1, 1) & -- & -- \\
\midrule
Proposal network \\
\dent Conv2D transpose && (4, 1) & (1, 1) & 128 & 128 \\
\dent Conv2D transpose && (5, 1) & (1, 1) & 128 & 64 \\
\bottomrule
\end{tabular}
\end{table}

\subsection{Post-processing hyperparameter selection}
\begin{table*}[t!]
\caption{Ranges of the hyperparameter values considered for (\subref{subtbl:search_params_threshold}) frame-level, and (\subref{subtbl:search_params_event}) event-level post-processing.}
    \label{tbl:search_params}
    \begin{subtable}{0.25\linewidth}
        \centering
        \caption{Frame-level hyperparameters.}
        \label{subtbl:search_params_threshold}
        \begin{tabular}{@{}lc@{}}
        \toprule
        Parameter & Search space \\
        \midrule
        Filter kernel           & [\,None, 11, 33, 55\,] \\
        On threshold            & $0.1\rightarrow0.9$ (inc. 0.1) \\
        Off threshold           & $0.1\rightarrow0.9$ (inc. 0.1) \\
        Hangover kernel  & [\,None, 11, 33, 55\,] \\
        \bottomrule
        \end{tabular}
    \end{subtable}
    \hspace{4em}
\begin{subtable}{0.60\linewidth}
        \renewcommand{\arraystretch}{1.2}
        \centering
        \caption{Event-level hyperparameters.}
        \label{subtbl:search_params_event}
        \begin{tabular}{@{}lccc@{}}
        \toprule
        Parameter                    & \texttt{bmabz}                   & \texttt{d}                       & \texttt{bp} \\
        \midrule
        Min. time between events~(s) & $0.1\rightarrow0.9$ (inc. 0.1)   & $0.1\rightarrow0.9$ (inc. 0.1)   & $0.1\rightarrow0.9$ (inc. 0.1) \\
        Min. event duration~(s)      & $2.0\rightarrow5.0$ (inc. 0.5)   & $0.6\rightarrow3.0$ (inc. 0.4)   & $0.3\rightarrow1.5$ (inc. 0.2)  \\
        Max. event duration~(s)      & $25.0\rightarrow40.0$ (inc. 2.5) & $5.0\rightarrow11.0$ (inc. 1.0)  & $2.0\rightarrow5.0$ (inc. 0.5)  \\
        \bottomrule
        \end{tabular}
    \end{subtable}
\end{table*}
During the evaluation of the different post-processing techniques, we employ a three-fold cross-validation scheme using the development sets~\autocite{stone1974crossval}.
For cross-validation, the data is partitioned into disjoint sets, referred to as \textit{folds}.
The development set consists of three site-year subsets, each of which is assigned to a different fold.
The best post-processing hyperparameters are chosen based on the highest F1-score over two of the folds~(development folds), while the third fold is held out for testing~(test fold).
After the parameters are chosen based on the development folds, the system is evaluated on the test fold.
The chosen development and test folds are permuted, referred to as a \textit{turn}, and the process is repeated.
The final system evaluation is computed by averaging the score over all three turns of the test folds.

\begin{figure}
    \centering

  \begin{subfigure}{\linewidth}
    \centering
    \caption{Overview of the post-processing process.}
    \label{fig:post-processing:overview}
    \vspace{0.75em} \includegraphics[width=0.95\linewidth]{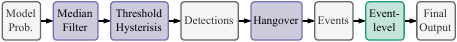}
  \end{subfigure}

  \begin{subfigure}{\linewidth}
    \centering
    \caption{Forward-search}
\label{fig:post-processing:forward}
    \includegraphics[width=0.95\linewidth]{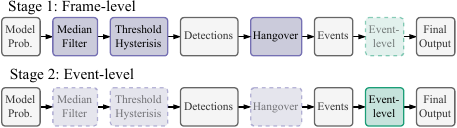}
  \end{subfigure}

  \begin{subfigure}{\linewidth}
    \centering
    \caption{Backward-search}
\label{fig:post-processing:backward}
    \includegraphics[width=0.95\linewidth]{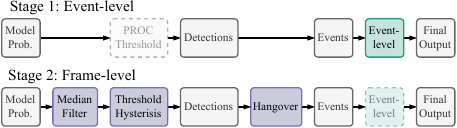}
  \end{subfigure}
  \caption{
  Illustration of the post-processing applied to the call probabilities produced by a per-frame \ac{sed} model.
(\subref{fig:post-processing:overview})
  An overview of the steps employed during both frame-level~(\textit{purple}) and event-level~(\textit{green}) post-processing.
(\subref{fig:post-processing:forward} and \subref{fig:post-processing:backward})
  The forward- and backward-search approaches to the selection of post-processing hyperparameters, respectively.
  Both optimisation methods consist of two stages, where some parts of the network remain fixed~(opaque dashes) while hyperparameters for the remainder are searched to maximise the F1-score over all development folds.
  }
  \label{fig:post-processing}
\end{figure}

Each post-processing hyperparameter is either applied at a frame-level or at an event-level, as discussed in \cref{subsec:background_post_processing}.
We propose two selection procedures, namely \textit{forward-search} and \textit{backward-search}.
During either search method, both frame-level and event-level hyperparameters are searched separately using a two-stage process.
An exhaustive optimisation across both hyperparameter types was practically infeasible, as it would require either substantial computational resources or a large reduction in the search space.

\subsubsection{Forward-search}
During the forward-search (refer to \cref{fig:post-processing:forward}), event-level hyperparameters are initially fixed (Stage 1) based on the statistical properties derived from the dataset.
The minimum inter-event duration is fixed at \qty{500}{ms}.
For each of the three target classes, the minimum and maximum event durations are drawn from the ranges in \Cref{tbl:call_info} by taking, within each group, the overall minimum and maximum of the constituent classes.
The frame-level hyperparameters are searched based on~\cref{subtbl:search_params_threshold}: median filter kernel size, threshold, hysteresis~(off) threshold, and the hangover kernel size.
Note that each of these hyperparameters is class-dependent.
Next, in Stage 2, these candidate frame-level hyperparameters are fixed, while the event-level hyperparameters are searched based on~\cref{subtbl:search_params_event}.

\subsubsection{Backward-search}
During the backward-search (refer to \cref{fig:post-processing:backward}), frame-level hyperparameters~(thresholds) are initially fixed (Stage 1) based on equal precision-recall, obtained from the average precision-recall curve computed over the development folds.
Event-level hyperparameters are searched based on~\cref{subtbl:search_params_event}.
Next, in Stage 2, these candidate event-level hyperparameters are fixed, while the frame-level hyperparameters are considered based on \cref{subtbl:search_params_threshold}.

\subsubsection{Final model evaluation}
After the hyperparameters have been fixed using either search method, precision and recall metrics are computed for each test fold.
The final F1-score, per call type, is then recomputed from the averaged precision/recall scores.
The final system evaluation is based on the macro F1-score of each class, again using the averaged precision/recall scores~\autocite{JeanLabadye2025BioDCASE2025Task2DevDataset}.
The hyperparameter selection process is repeated for both the \gls{wad} and the \gls{wadbpn} models.

\section{Results}

\Cref{tbl:results_cross_val} shows the cross-validated performance of the baseline model when applying hyperparameter selection, using backward-search, on a per-class basis.
Applying event-level optimisation alone led to consistent performance gains for all classes, when compared to the previously-used empirical method of selecting the post-processing parameters (\cref{tbl:results_cross_val_no_processing}).
In particular, for D-calls~(\texttt{d}) we see a \qty{9.3}{\percent} absolute improvement in the F1-score, which corresponds to a \qty{77.5}{\percent} relative improvement over the empirically derived parameters.
Additional frame-level optimisation provides more modest gains, leading to a final macro F1-score of \num{0.422} across all classes.
Overall, this represents a \qty{4.5}{\percent} absolute improvement in the macro F1-score, compared to the baseline \gls{wad} model, achieved solely through the optimisation of post-processing and no architectural modifications (see \cref{tbl:final_model_results}).
Forward-search achieved similar performance to backward-search.
However, since backward-search performed slightly better, results for only this method will be shown.

\begin{figure}[t]
    \centering
    \includegraphics[width=\linewidth]{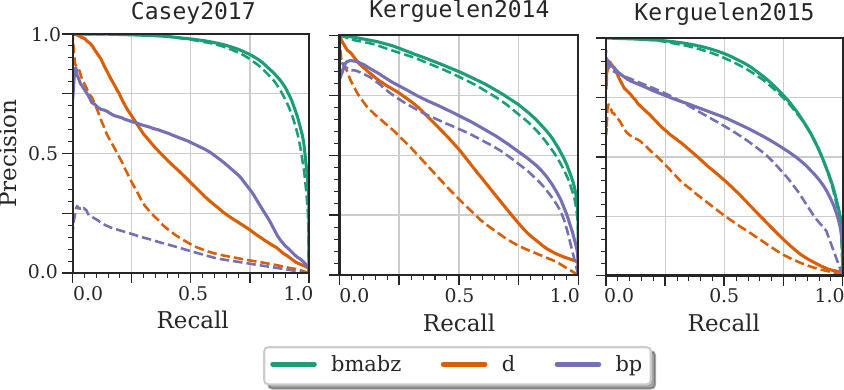}
    \caption{
    Precision-recall curves for each call type and for each of the three site-year development sets.
    Both the baseline \gls{wad}[~(dashed)] and the proposed \gls{wadbpn}[~(solid)] models are shown.
    }
    \label{fig:proc_curves}
\end{figure}

\Cref{fig:proc_curves} shows that the \gls{wadbpn} architecture yields superior precision-recall curves across all three call types relative to the original \gls{wad} system, indicating better performance even without optimisation of the post-processing.
The model successfully leads to improved precision~(fewer false positives) across a broad range of classification thresholds and particularly for the minority classes \texttt{d} and \texttt{bp}.

The application of hyperparameter optimisation using backward-search for the proposed \gls{wadbpn} model is shown in \Cref{tbl:results_cross_val}.
Frame-level optimisation achieves substantial improvements and is more successful than event-level optimisation.
Specifically, \texttt{bmabz}-, \texttt{d}- and \texttt{bp}-calls improve by \qty{6.9}{\percent}, \qty{9.8}{\percent}, and \qty{4.8}{\percent}, respectively.
The final macro F1-score reaches \num{0.475}, representing a \qty{9.8}{\percent} absolute improvement over the original \gls{wad} model and a \qty{5.3}{\percent} improvement over \gls{wad} with optimised post-processing (see \cref{tbl:final_model_results}).

To investigate the individual contribution of each frame-level post-processing technique, we also evaluated the absolute improvement when each technique is applied in addition to selecting class-dependent thresholds.
Specifically, we consider class-dependent hysteresis thresholds, hangover, and median filtering.
Although a few cases yielded noticeable gains, most contributed relatively little compared to class-dependent threshold optimisation alone.
As shown in \cref{tbl:results_cross_val}, event-level hyperparameter optimisation produces large performance gains, challenging the prevalent current practice of simply deriving these values from dataset statistics.

Finally, \cref{tbl:final_model_results} highlights the trade-offs underlying these gains.
The optimised \gls{wad} model improves the F1-score primarily by sacrificing recall in exchange for higher precision.
By contrast, \gls{wadbpn} manages to achieve a marginal improvement in recall compared to the baseline \gls{wad} model whilst also achieving a substantial increase in precision.
Therefore, the boundary proposal network has succeeded in reducing false positives while preserving recall.

\begin{table*}[t!]
    \centering
    \caption{
    Cross-validation F1-score for \gls{wad}~(baseline) and \gls{wadbpn} models using a backward-search for selecting the post-processing hyperparameters.
(\subref{subtbl:results_cross_val_bpn_event}, Stage 1)
Event-level: classification thresholds are selected from the average precision-recall curve over the development folds, after which event-level hyperparameters are selected.
(\subref{subtbl:results_cross_val_bpn_frame}, Stage 2)
Frame-level: hyperparameters are selected, while event-level hyperparameters remain fixed at the previously selected values.
The final F1-score is recalculated based on the average of the precision and recall for each test fold.
}
    \label{tbl:results_cross_val}
    \begin{subtable}{0.48\linewidth}
        \centering
        \caption{Stage 1: Event-level selection}
        \label{subtbl:results_cross_val_event}
        \label{subtbl:results_cross_val_bpn_event}
        \begin{tabular}{@{}lccccccc@{}}
            \toprule
            \multirow{2}{*}{Call type} & \multicolumn{2}{c}{Fold 1} & \multicolumn{2}{c}{Fold 2} & \multicolumn{2}{c}{Fold 3} & \multirow{2}{*}{Final F1} \\
                                       & Dev          & Test        & Dev          & Test        & Dev          & Test        &                           \\ \midrule
        \multicolumn{8}{c}{\Gls{wad}}                                                                                                                     \\ \midrule
            \texttt{bmabz}             & 0.634        & 0.686       & 0.665        & 0.620       & 0.670        & 0.591       & \textbf{0.663}            \\
            \texttt{d}                 & 0.212        & 0.162       & 0.226        & 0.186       & 0.199        & 0.220       & 0.219                     \\
            \texttt{bp}                & 0.529        & 0.029       & 0.291        & 0.446       & 0.237        & 0.560       & 0.348                     \\ \midrule
        \multicolumn{8}{c}{\Gls{wadbpn}}                                                                                                                  \\ \midrule
            \texttt{bmabz}             & 0.546        & 0.510       & 0.536        & 0.644       & 0.526        & 0.410       & 0.546                     \\
            \texttt{d}                 & 0.238        & 0.233       & 0.204        & 0.152       & 0.235        & 0.307       & \textbf{0.242}            \\
            \texttt{bp}                & 0.512        & 0.288       & 0.376        & 0.451       & 0.339        & 0.489       & \textbf{0.412}            \\ \bottomrule
        \end{tabular}
    \end{subtable}
    \hspace{\columnsep}
    \begin{subtable}{0.48\linewidth}
        \centering
        \caption{Stage 2: Frame-level selection}
        \label{subtbl:results_cross_val_frame}
        \label{subtbl:results_cross_val_bpn_frame}
        \begin{tabular}{@{}lccccccc@{}}
            \toprule
            \multirow{2}{*}{Call type} & \multicolumn{2}{c}{Fold 1} & \multicolumn{2}{c}{Fold 2} & \multicolumn{2}{c}{Fold 3} & \multirow{2}{*}{Final F1} \\
                                       &  Dev          & Test       & Dev          & Test        & Dev          & Test        &                           \\ \midrule
        \multicolumn{8}{c}{\Gls{wad}}                                                                                                                     \\ \midrule
            \texttt{bmabz}             &  0.626        & 0.698       & 0.659        & 0.632       & 0.665        & 0.621       & \textbf{0.669}           \\
            \texttt{d}                 &  0.209        & 0.223       & 0.233        & 0.175       & 0.199        & 0.242       & 0.222                    \\
            \texttt{bp}                &  0.530        & 0.029       & 0.302        & 0.485       & 0.257        & 0.575       & 0.363                    \\ \midrule
        \multicolumn{8}{c}{\Gls{wadbpn}}                                                                                                                  \\ \midrule
            \texttt{bmabz}             &  0.590        & 0.603       & 0.585        & 0.614       & 0.608        & 0.567       & 0.615                    \\
            \texttt{d}                 &  0.311        & 0.388       & 0.377        & 0.256       & 0.322        & 0.366       & \textbf{0.340}           \\
            \texttt{bp}                &  0.532        & 0.311       & 0.442        & 0.492       & 0.401        & 0.573       & \textbf{0.460}           \\ \bottomrule
        \end{tabular}
    \end{subtable}
\end{table*}

\begin{table}[t]
    \centering
    \caption{Cross-validation F1-score for \gls{wad} when no frame-level post-processing is applied and event-level hyperparameters are selected from empirical call statistics.}
    \label{tbl:results_cross_val_no_processing}
    \begin{tabular}{@{}lccccccc@{}}
    \toprule
    \multirow{2}{*}{Type} & \multicolumn{2}{c}{Fold 1} & \multicolumn{2}{c}{Fold 2} & \multicolumn{2}{c}{Fold 3} & \multirow{2}{*}{Final F1} \\
                & Dev & Test & Dev & Test & Dev & Test & \\
    \midrule
    \texttt{bmabz}     & 0.615 & 0.620 & 0.620 & 0.657 & 0.631 & 0.536 & 0.628 \\
    \texttt{d}         & 0.136 & 0.049 & 0.106 & 0.128 & 0.099 & 0.168 & 0.126 \\
    \texttt{bp}        & 0.494 & 0.038 & 0.286 & 0.386 & 0.217 & 0.521 & 0.315 \\
    \bottomrule
    \end{tabular}
\end{table}

\begin{table}[t]
    \centering
    \caption{
    Final results comparing the original \gls{wad} architecture, a variant with post-processing optimised via cross-validation, and the best reported model, \gls{wadbpn}, also with optimisation. Scores are averaged across call types and test folds.
    }
    \label{tbl:final_model_results}
    \begin{tabular}{lcccc}
    \toprule
    Model                                       & \acs{iou} & Recall & Precision & F1 \\
    \midrule
    WhaleVAD~\autocite{Geldenhuys2025WhaleVAD}  & 0.588     & 0.458  & 0.320     & 0.377 \\
    WhaleVAD + optimisation                     & 0.601     & 0.391  & 0.458     & 0.422 \\
WhaleVAD-BPN + optimisation                 & \textbf{0.625}     & \textbf{0.463}  & \textbf{0.488}     & \textbf{0.475} \\
    \bottomrule
    \end{tabular}
\end{table}

\section{Conclusion}
In this work, we present a novel and computationally lightweight network augmentation~(\ac{bpn}) for an existing whale call detection system, as well as a computationally tractable approach to hyperparameter selection for the system post-processing.
Both innovations are shown to lead to substantial performance improvements over the baseline system using the 2025 DCASE~(Task 2) challenge data, for automated baleen whale call detection.

By exploiting intermediate latent features already computed within the main classifier to act as a gating mechanism for the output, the \ac{bpn} consistently reduces false positive rates across all classes.
The augmented architectures have also shown to improve minority-class call detection, which is generally more difficult than the detection of abundant classes.
This is important because annotated data is difficult to obtain, and therefore better performance for a small pool of training examples is highly desirable.

We further demonstrate that the principled selection of post-processing hyperparameters has a marked impact on final system performance.
We compare two hyperparameter selection strategies, namely a forward- and backward-search, which both achieve comparable gains.
When comparing hyperparameters selected in this way to conventional empirical or ad-hoc choices, a \qty{4.5}{\percent} absolute improvement is seen.

Our final system, which includes the proposed \gls{bpn} and optimised post-processing hyperparameters, achieves a \qty{9.8}{\percent} absolute improvement in overall F1-score.
The system succeeds in markedly reducing false positives, while improving the detection of minority-class calls.
In addition to maintaining the already strong recall performance of the baseline baleen whale call detection system, these improvements narrow the performance gap between minority-class calls and calls for which there is abundant data.
The proposed system should therefore be a useful tool for the discovery and monitoring of new call types, for which data will initially always be limited.

\section*{Acknowledge}
\noindent
The authors gratefully acknowledges Telkom (South Africa) for their financial support and thanks the Stellenbosch Rhasatsha high performance computing~(HPC1) facility for the compute time provided to the research presented in this paper.
The authors also acknowledges the anonymous reviewers for their insightful comments, which have improved the clarity of this manuscript.

\printbibliography
\end{document}

%% file: glossary.tex
\newacronym{aru}{ARU}{automatic recording unit}
\newacronym{ssl}{SSL}{self-supervised learning}
\newacronym{bilstm}{BiLSTM}{bidirectional long short-term memory network}
\newacronym{ast}{AST}{audio spectrogram transformer}
\newacronym{vit}{ViT}{vision transformer}
\newacronym{auc}{AUC}{area under the curve}
\newacronym{fpr}{FPR}{false positive rate}
\newacronym{tpr}{TPR}{true positive rate}
\newacronym{auc roc}{AUC ROC}{area under the receiver operating characteristic curve}
\newacronym{roc}{ROC}{receiver operating characteristic}
\newacronym{stft}{STFT}{short time Fourier transform}
\newacronym{fft}{FFT}{fast Fourier transform}
\newacronym{nlp}{NLP}{natural language processing}
\newacronym{cnn}{CNN}{convolutional neural network}
\newacronym{ldc}{LDC}{Linguistic Data Consortium}
\newacronym{mfcc}{MFCC}{mel frequency cepstral coefficient}
\newacronym{lfcc}{LFCC}{linear frequency cepstral coefficient}
\newacronym{knn}{k-NN}{k-nearest neighbours}
\newacronym{elev}{EV}{Elephant Voices}
\newacronym{elp}{ELP}{Elephant Listening Project}
\newacronym{dt}{DT}{decision tree}
\newacronym{rf}{RF}{random forest}
\newacronym{lr}{LR}{logistic regression}
\newacronym{mlp}{MLP}{multi-layer perceptron}
\newacronym{svm}{SVM}{support vector machine}
\newacronym{xgb}{XGB}{XGBoost}
\newacronym{dft}{DFT}{discrete Fourier transform}
\newacronym{lda}{LDA}{linear discriminant analysis}
\newacronym{lpc}{LPC}{linear predictive coding}
\newacronym{hmm}{HMM}{hidden Markov model}
\newacronym{pam}{PAM}{passive acoustic monitoring}
\newacronym{rnn}{RNN}{recurrent neural network}
\newacronym{map}{mAP}{mean average precision}
\newacronym{ap}{AP}{average precision}
\newacronym{pca}{PCA}{principal component analysis}
\newacronym{resnet}{ResNet}{residual neural network}
\newacronym{iou}{IoU}{intersection over union}
\newacronym{beats}{BEATs}{bidirectional encoder representation from audio transformers}
\newacronym{iucn}{IUCN}{International Union for Conservation of Nature}
\newacronym{rbf}{RBF}{radial basis function}
\newacronym{umap}{UMAP}{uniform manifold approximation and projection}
\newacronym{gpu}{GPU}{graphics processing unit}
\newacronym{lmu}{LMU}{Legendre memory unit}
\newacronym{dec}{DEC}{deep embedded clustering}
\newacronym{nmi}{NMI}{normalised mutual information}
\newacronym{drl}{DRL}{disentangled representation learning}
\newacronym{vae}{VAE}{variational autoencoder}
\newacronym{lstm}{LSTM}{long short-term memory}
\newacronym{gru}{GRU}{gated recurrent unit}
\newacronym{tsne}{t-SNE}{t-distributed stochastic neighbour embedding}
\newacronym{gmm}{GMM}{Gaussian mixture model}
\newacronym{clap}{CLAP}{contrastive language-audio pretraining}
\newacronym{asr}{ASR}{automatic speech recognition}
\newacronym{mds}{MDS}{multidimensional scaling}
\newacronym{dtw}{DTW}{dynamic time warping}
\newacronym{aerd}{AERD}{automatic elephant rumble detection}
\newacronym{dct}{DCT}{discrete cosine transform}
\newacronym{sanctsound}{Sanctsound}{Sanctuary Soundscape Monitoring Project}
\newacronym{ieee}{IEEE}{Institute of Electrical and Electronics Engineers}
\newacronym{sed}{SED}{sound event detection}
\newacronym{snr}{SNR}{signal-to-noise ratio}
\newacronym{ann}{ANN}{artificial neural network}
\newacronym{crnn}{CRNN}{convolutional recurrent neural network}
\newacronym{atbfl}{ATBFL}{Acoustic Trends Blue Fin Library}
\newacronym{atp}{ATP}{Antarctic Blue and Fin Whale Acoustic Trends Project}
\newacronym{iwc-sorp}{IWC-SORP}{International Whaling Commission's Southern Ocean Research Partnership}
\newacronym{noaa}{NOAA}{National Oceanic and Atmospheric Administration}
\newacronym{bce}{BCE}{binary cross-entropy}
\newacronym{tp}{TP}{true positive}
\newacronym{fp}{FP}{false positive}
\newacronym{fn}{FN}{false negative}
\newacronym{vad}{VAD}{voice activity detection}
\newacronym{dcase}{DCASE}{Detection and Classification of Acoustic Scenes and Events}
\newacronym{cnn-bilstm}{CNN-BiLSTM}{convolutional bidirectional long short-term memory neural network}
\newacronym{nms}{NMS}{non-maximum suppression}
\newacronym{bpn}{BPN}{boundary proposal network}
\newacronym{rpn}{RPN}{region proposal network}
\newacronym{fpn}{FPN}{feature pyramid network}
\newacronym[longplural={regions of interest}]{roi}{ROI}{region of interest}
\newacronym{dnn}{DNN}{deep neural network}

\newglossaryentry{ava-vad}{
	name={AVA-VAD},
	description={},
}

\newglossaryentry{wadbpn}{
	name={WhaleVAD-BPN},
	description={},
}

\newglossaryentry{wad}{
	name={WhaleVAD},
	description={},
}

\newglossaryentry{librispeech-960h}{
	name={LibriSpeech ASR~(960h)},
	description={},
}

\newglossaryentry{beans20}{
	name={BEANs-20},
	description={},
}

\newglossaryentry{beans40}{
	name={BEANs-40},
	description={},
}

\newglossaryentry{vggish}{
	name={VGGish},
	description={},
}

\newglossaryentry{vgg}{
	name={VGG},
	description={},
}

\newglossaryentry{birdnet}{
	name={BirdNet},
	description={},
}

\newglossaryentry{perch}{
	name={Perch},
	description={},
}

\newglossaryentry{perch8}{
	name={Perch (ver. 8)},
	description={},
}

\newglossaryentry{surfperch}{
	name={SurfPerch},
	description={},
}

\newglossaryentry{sierras}{
	name={Sierras},
	description={},
}

\newglossaryentry{psla}{
	name={PSLA},
	description={},
}

\newglossaryentry{yamnet}{
	name={YAMNet},
	description={},
}

\newglossaryentry{audiomae}{
	name={AudioMAE},
	description={},
}

\newglossaryentry{hubert}{
	name={HuBERT},
	description={},
}

\newglossaryentry{hubert-base}{
	name={HuBERT~(base)},
	description={},
}

\newglossaryentry{hubert-large}{
	name={HuBERT~(large)},
	description={},
}

\newglossaryentry{hubert-xlarge}{
	name={HuBERT~(xlarge)},
	description={},
}

\newglossaryentry{wav2vec2}{
	name={wav2vec2},
	description={},
}

\newglossaryentry{w2v2-base}{
	name={wav2vec2~(base)},
	description={},
}

\newglossaryentry{w2v2-large}{
	name={wav2vec2~(large)},
	description={},
}

\newglossaryentry{xlsr}{
	name={XLSR},
	description={},
}

\newglossaryentry{multi-whale}{
	name={multi-species whale},
	description={},
}

\newglossaryentry{humpback}{
	name={humpback},
	description={},
}

\newglossaryentry{aves}{
	name={AVES},
	description={},
}

\newglossaryentry{aves-core}{
	name={AVES-core},
	description={},
}

\newglossaryentry{aves-bio}{
	name={AVES-bio},
	description={},
}

\newglossaryentry{aves-nonbio}{
	name={AVES-nonbio},
	description={},
}

\newglossaryentry{aves-all}{
	name={AVES-all},
	description={},
}

\newglossaryentry{birdaves}{
	name={BirdAVES},
	description={},
}

\newglossaryentry{birdaves-biox-base}{
	name={BirdAVES-biox-base},
	description={},
}

\newglossaryentry{birdaves-biox-large}{
	name={BirdAVES-biox-large},
	description={},
}

\newglossaryentry{birdaves-bioxn-large}{
	name={BirdAVES-bioxn-large},
	description={},
}

\newglossaryentry{xenocanto}{
	name={Xeno-canto},
	description={},
}

\newglossaryentry{inaturalist}{
	name={iNaturalist},
	description={},
}

\newglossaryentry{efficientnet-b1}{
	name={EfficientNet-B1},
	description={},
}

\newglossaryentry{birdnetv2.4}{
	name={BirdNET~(ver. 2.4)},
	description={},
}

\newglossaryentry{fsd50k}{
	name={FSD50k},
	description={},
}

\newglossaryentry{audioset}{
	name={AudioSet},
	description={},
}

\newglossaryentry{vggsound}{
	name={VGGSound},
	description={},
}

\newglossaryentry{hdbscan}{
	name={HDBSCAN},
	description={},
}

\newglossaryentry{chatgpt}{
	name={ChatGPT},
	description={},
}

\newglossaryentry{ast-seq}{
	name={AST-seq},
	description={},
}

\newglossaryentry{ast-lab}{
	name={AST-cls},
	description={},
}

\newglossaryentry{lr-frame}{
	name={LR-frame},
	description={},
}

%% file: references.bib
@article{miller2021atbfljournal,
	title = {An open access dataset for developing automated detectors of Antarctic baleen whale sounds and performance evaluation of two commonly used detectors},
	volume = {11},
	issn = {2045-2322},
	doi = {10.1038/s41598-020-78995-8},
	shorttitle = {{ATBFL}},
	pages = {806},
	number = {1},
	journaltitle = {Scientific Reports},
	shortjournal = {Sci Rep},
	author = {Miller, Brian S. and {The IWC-SORP/SOOS Acoustic Trends Working Group} and Stafford, Kathleen M. and Van Opzeeland, Ilse and Harris, Danielle and Samaran, Flore and Širović, Ana and Buchan, Susannah and Findlay, Ken and Balcazar, Naysa and Nieukirk, Sharon and Leroy, Emmanuelle C. and Aulich, Meghan and Shabangu, Fannie W. and Dziak, Robert P. and Lee, Won Sang and Hong, Jong Kuk},
	date = {2021},
}

@misc{Liu2022SurreySF,
	title = {Surrey system for {DCASE} 2022 task 5: Few-shot bioacoustic event detection with segment-level metric learning},
	doi = {10.48550/arXiv.2207.10547},
	number = {{arXiv}:2207.10547v1},
	publisher = {{arXiv}},
	author = {Liu, Haohe and Liu, Xubo and Mei, Xinhao and Kong, Qiuqiang and Wang, Wenwu and Plumbley, Mark D.},
	date = {2022},
}

@misc{Zhao2024FewShotBE,
	title = {Few-shot bioacoustic event detection with frame-level embedding learning system},
	doi = {10.48550/arXiv.2407.10182},
	number = {{arXiv}:2407.10182},
	publisher = {{arXiv}},
	author = {Zhao, {PengYuan} and Lu, {ChengWei} and Zou, Li-Wei},
	date = {2024},
}

@report{hoffman2024DCASE,
	title = {Toward in-context bioacoustic sound event detection},
	institution = {{DCASE}2024 {CHallenge}},
	type = {Technical Report},
	author = {Hoffman, Benjamin and Robinson, David},
	date = {2024},
}

@report{Geldenhuys2025WhaleVAD,
	title = {Whale-{VAD}: Whale Vocalisation Activity Detection},
	institution = {{DCASE}2025 Challenge},
	type = {Technical Report},
	author = {Geldenhuys, Christiaan M and Tonitz, Günther and Niesler, Thomas R},
	date = {2025},
}

@misc{Lin2018FocalLoss,
	title = {Focal Loss for Dense Object Detection},
	doi = {10.48550/arXiv.1708.02002},
	number = {{arXiv}:1708.02002},
	publisher = {{arXiv}},
	author = {Lin, Tsung-Yi and Goyal, Priya and Girshick, Ross and He, Kaiming and Dollár, Piotr},
	date = {2018},
	eprinttype = {arxiv},
	eprint = {1708.02002},
}

@inproceedings{losh2017adamw,
	location = {Toulon, France},
	title = {Decoupled weight decay regularization},
	booktitle = {Proceedings of International Conference on Learning Representations ({ICLR})},
	author = {Loshchilov, I. and Hutter, F.},
	date = {2017},
}

@article{Rasmussen2021AutomaticDA,
	title = {Automatic detection and classification of baleen whale social calls using convolutional neural networks.},
	volume = {149},
	pages = {3635},
	number = {5},
	journaltitle = {The Journal of the Acoustical Society of America},
	author = {Rasmussen, Jeppe Have and Sirovic, Ana},
	date = {2021},
}

@article{Miller2023dcalls,
	title = {Deep learning algorithm outperforms experienced human observer at detection of blue whale D‐calls: a double‐observer analysis},
	volume = {9},
	doi = {10.1002/rse2.297},
	pages = {104--116},
	number = {1},
	journaltitle = {Remote Sensing in Ecology and Conservation},
	shortjournal = {Remote Sens Ecol Conserv},
	author = {Miller, Brian S. and Madhusudhana, Shyam and Aulich, Meghan G. and Kelly, Nat},
	editor = {Lecours, Vincent and Risch, Denise},
	date = {2023},
}

@misc{Huang2018DenseNet,
	title = {Densely Connected Convolutional Networks},
	doi = {10.48550/arXiv.1608.06993},
	number = {{arXiv}:1608.06993},
	publisher = {{arXiv}},
	author = {Huang, Gao and Liu, Zhuang and Maaten, Laurens van der and Weinberger, Kilian Q.},
	date = {2018},
}

@inproceedings{Ibrahim2016SVMNorth,
	title = {A New Approach for North Atlantic Right Whale Upcall Detection},
	doi = {10.1109/IS3C.2016.76},
	pages = {260--263},
	booktitle = {2016 International Symposium on Computer, Consumer and Control ({IS}3C)},
	author = {Ibrahim, Ali K. and Zhuang, Hanqi and Erdol, Nurgun and Ali, Ali Muhamed},
	date = {2016},
}

@inproceedings{Cuevas2017GMMBlue,
	title = {Unsupervised blue whale call detection using multiple time-frequency features},
	doi = {10.1109/CHILECON.2017.8229663},
	pages = {1--6},
	booktitle = {2017 {CHILEAN} Conference on Electrical, Electronics Engineering, Information and Communication Technologies ({CHILECON})},
	author = {Cuevas, Alejandro and Veragua, Alejandro and Español-Jiménez, Sonia and Chiang, Gustavo and Tobar, Felipe A.},
	date = {2017},
}

@inproceedings{Mazhar2007SVMHumpback,
	title = {Vocalization based Individual Classification of Humpback Whales using Support Vector Machine},
	doi = {10.1109/OCEANS.2007.4449356},
	pages = {1--9},
	booktitle = {{OCEANS} 2007},
	author = {Mazhar, Suleman and Ura, Tamaki and Bahl, Rajendar},
	date = {2007},
}

@inproceedings{pace2012HMMHumpback,
	title = {Hidden Markov Modeling for humpback whale (Megaptera Novaeanglie) call classification},
	volume = {17},
	doi = {10.1121/1.4772751},
	eventtitle = {Acoustical Society of America},
	booktitle = {Meetings on Acoustics},
	author = {Pace, Federica and White, Paul R. and Adam, Olivier},
	date = {2012},
}

@article{Mellinger2000SpecCor,
	title = {Recognizing transient low-frequency whale sounds by spectrogram correlation.},
	volume = {107},
	doi = {10.1121/1.429434},
	pages = {3518--29},
	number = {6},
	journaltitle = {The Journal of the Acoustical Society of America},
	author = {Mellinger, David K. and Clark, Chris},
	date = {2000},
}

@article{Mellinger1997MysticeteMethods,
	title = {Methods for automatic detection of mysticete sounds},
	volume = {29},
	pages = {163--181},
	journaltitle = {Marine and Freshwater Behaviour and Physiology},
	author = {Mellinger, David K. and Clark, Christopher W.},
	date = {1997},
}

@inproceedings{Luo2016RecpetiveFieldCNN,
	location = {Barcelona, Spain},
	title = {Understanding the Effective Receptive Field in Deep Convolutional Neural Networks},
	volume = {29},
	eventtitle = {Advances in Neural Information Processing Systems},
	booktitle = {Proceedings of Advances in Neural Information Processing Systems ({NIPS} 2016)},
	publisher = {Curran Associates, Inc.},
	author = {Luo, Wenjie and Li, Yujia and Urtasun, Raquel and Zemel, Richard},
	editor = {Lee, D. and Sugiyama, M. and Luxburg, U. and Guyon, I. and Garnett, R.},
	date = {2016},
}

@article{stone1974crossval,
	title = {Cross-validatory choice and assessment of statistical predictions},
	volume = {36},
	series = {B},
	pages = {111--133},
	number = {2},
	journaltitle = {Journal of the Royal Statistical Society},
	author = {Stone, Mervyn},
	date = {1974},
}

@inproceedings{Lin2017FeaturePyramidNetwork,
	location = {Las Vegas, {USA}},
	title = {Feature Pyramid Networks for Object Detection},
	eventtitle = {Computer Vision and Pattern Recognition ({CVPR})},
	pages = {2117--2125},
	booktitle = {Proceedings of the {IEEE} conference on Computer Vision and Pattern Recognition ({CVPR})},
	author = {Lin, Tsung-Yi and Dollár, Piotr and Girshick, Ross and He, Kaiming and Hariharan, Bharath and Belongie, Serge},
	date = {2017},
}

@misc{Redmon2018YOLOv3,
	title = {{YOLOv}3: An Incremental Improvement},
	doi = {10.48550/arXiv.1804.02767},
	shorttitle = {{YOLOv}3},
	number = {{arXiv}:1804.02767},
	publisher = {{arXiv}},
	author = {Redmon, Joseph and Farhadi, Ali},
	date = {2018},
	eprinttype = {arxiv},
	eprint = {1804.02767 [cs]},
}

@inproceedings{Liu2016SSD,
	location = {Amsterdam, Netherlands},
	title = {{SSD}: Single shot multibox detector},
	eventtitle = {European Conference on Computer Vision},
	pages = {21--37},
	booktitle = {Proceedings of the European Conference on Computer Vision ({ECCV})},
	publisher = {Springer},
	author = {Liu, Wei and Anguelov, Dragomir and Erhan, Dumitru and Szegedy, Christian and Reed, Scott and Fu, Cheng-Yang and Berg, Alexander C},
	date = {2016},
}

@inproceedings{Girshick2014RCNN,
	title = {Rich feature hierarchies for accurate object detection and semantic segmentation},
	pages = {580--587},
	booktitle = {Proceedings of the {IEEE} conference on computer vision and pattern recognition},
	author = {Girshick, Ross and Donahue, Jeff and Darrell, Trevor and Malik, Jitendra},
	date = {2014},
}

@inproceedings{Girshick2015FastRCNN,
	title = {Fast R-{CNN}},
	pages = {1440--1448},
	booktitle = {Proceedings of the {IEEE} international conference on computer vision},
	author = {Girshick, Ross},
	date = {2015},
}

@article{Uijlings2013SelectiveSearch,
	title = {Selective Search for Object Recognition},
	volume = {104},
	issn = {1573-1405},
	doi = {10.1007/s11263-013-0620-5},
	pages = {154--171},
	number = {2},
	journaltitle = {International Journal of Computer Vision},
	shortjournal = {Int J Comput Vis},
	author = {Uijlings, J. R. R. and van de Sande, K. E. A. and Gevers, T. and Smeulders, A. W. M.},
	date = {2013},
}

@article{Kowarski2021BaleenReview,
	title = {A review of big data analysis methods for baleen whale passive acoustic monitoring},
	volume = {37},
	issn = {0824-0469, 1748-7692},
	doi = {10.1111/mms.12758},
	pages = {652--673},
	number = {2},
	journaltitle = {Marine Mammal Science},
	shortjournal = {Marine Mammal Science},
	author = {Kowarski, Katie A. and Moors‐Murphy, Hilary},
	date = {2021},
}

@article{Cooke2018BlueWhale,
	title = {Balaenoptera musculus},
	journaltitle = {The {IUCN} Red List of Threatened Species},
	author = {Cooke, J. G.},
	date = {2018},
	note = {Erratum published in 2019},
}

@article{Cooke2018FinWhale,
	title = {Balaenoptera physalus},
	journaltitle = {The {IUCN} Red List of Threatened Species},
	author = {Cooke, J. G.},
	date = {2018},
}

@inproceedings{Ren2015FasterRCNN,
	location = {Montreal, Canada},
	title = {Faster R-{CNN}: Towards Real-Time Object Detection with Region Proposal Networks},
	volume = {28},
	booktitle = {Proceedings of Advances in Neural Information Processing Systems ({NIPS} 2015)},
	author = {Ren, Shaoqing and He, Kaiming and Girshick, Ross and Sun, Jian},
	date = {2015},
}

@inproceedings{Thompson2015SpatialDropout,
	title = {Efficient object localization using convolutional networks},
	pages = {648--656},
	booktitle = {Proceedings of the {IEEE} conference on computer vision and pattern recognition},
	author = {Tompson, Jonathan and Goroshin, Ross and Jain, Arjun and {LeCun}, Yann and Bregler, Christoph},
	date = {2015},
}

@report{Cances2019MULTITASKLA,
	title = {Multi-task learning and post processing optimization for sound event detection},
	institution = {{DCASE}2019 Challenge},
	type = {Technical Report},
	author = {Cances, Léo and Pellegrini, Thomas and Guyot, Patrice},
	date = {2019},
}

@article{Barshan1998PerformanceCO,
	title = {Performance comparison of four time-of-flight estimation methods for sonar signals},
	volume = {34},
	doi = {10.1049/el:19981127},
	pages = {1616--1617},
	number = {16},
	journaltitle = {Electronics Letters},
	author = {Barshan, Billur and Ayrulu, Birsel},
	date = {1998},
}

@book{difranco1980radar,
	title = {Radar detection},
	isbn = {978-0-89006-092-6},
	series = {The Artech radar library},
	publisher = {Artech House},
	author = {{DiFranco}, J.V. and Rubin, W.L.},
	date = {1980},
}

@misc{Cances2019EvalPostprocessingDCASE,
	title = {Evaluation of post-processing algorithms for polyphonic sound event detection},
	doi = {10.48550/arXiv.1906.06909},
	number = {{arXiv}:1906.06909},
	publisher = {{arXiv}},
	author = {Cances, Leo and Guyot, Patrice and Pellegrini, Thomas},
	date = {2019},
}

@article{Schall2024Baleen,
	title = {Deep learning in marine bioacoustics: a benchmark for baleen whale detection},
	volume = {10},
	issn = {2056-3485, 2056-3485},
	doi = {10.1002/rse2.392},
	shorttitle = {Deep learning in marine bioacoustics},
	pages = {642--654},
	number = {5},
	journaltitle = {Remote Sensing in Ecology and Conservation},
	author = {Schall, Elena and Kaya, Idil Ilgaz and Debusschere, Elisabeth and Devos, Paul and Parcerisas, Clea},
	date = {2024},
}

@inproceedings{Wilkinson2021AVAVAD,
	location = {Toronto, Canada},
	title = {A Hybrid {CNN}-{BiLSTM} Voice Activity Detector},
	isbn = {978-1-72817-605-5},
	doi = {10.1109/ICASSP39728.2021.9415081},
	pages = {6803--6807},
	booktitle = {{IEEE} International Conference on Acoustics, Speech and Signal Processing ({ICASSP})},
	publisher = {{IEEE}},
	author = {Wilkinson, Nicholas and Niesler, Thomas},
	date = {2021},
}


%% file: references_datasets.bib
@dataset{Miller2020ATBFL,
	title = {An annotated library of underwater acoustic recordings for testing and training automated algorithms for detecting Antarctic blue and fin whale sounds},
	doi = {10.26179/5e6056035c01b},
	version = {1},
	publisher = {Australian Antarctic Data Centre},
	author = {Miller, Brian Seth and Stafford, Kathleen M and Van Opzeeland, Ilse and Harris, Danielle and Samaran, Flore and Š{IROVIĆ}, {ANA} and Buchan, Susannah and Findlay, Ken and Balcazar, Naysa and Nieukirk, Sharon and Leroy, Emmanuelle C and Aulich, Meghan and Shabangu, Fannie W and Dziak, Robert P and Lee, Won Sang and Hong, Jong Kuk},
	date = {2020},
}

@dataset{JeanLabadye2025BioDCASE2025Task2DevDataset,
	title = {{BioDCASE} 2025 Task 2: Development set},
	doi = {10.5281/zenodo.15092732},
	version = {1 (Erratum)},
	publisher = {Zenodo},
	author = {Jean-Labadye, Lucie and Parcerisas, Clea and Miller, Brian and Carvaillo, Paul and Dubus, Gabriel and Farrugia, Nicolas and Gros-Martial, Anatole and Marmoret, Axel and Moummad, Ilyass and {Andrea Napoli} and Nguyen Hong Duc, Paul and Raumer, Pierre-Yves and Schall, Elena and White, Ellen and Adam, Olivier and Roch, Marie A. and White, Paul and Cazau, Dorian},
	date = {2025},
}
